\documentclass[journal, twoside]{IEEEtran}
\usepackage[utf8]{inputenc}
\usepackage[per-mode=symbol]{siunitx}
\DeclareSIUnit{\sample}{Sa}
\DeclareSIUnit{\baud}{Bd}
\DeclareSIUnit{\bit}{b}
\DeclareSIUnit{\byte}{B}

\usepackage{hyperref}
\hypersetup{hidelinks}

\usepackage[acronym,nomain]{glossaries}
\glsdisablehyper

\usepackage{lipsum}
\usepackage{graphicx}
\usepackage{grffile}
\usepackage{booktabs}
\usepackage{nicefrac}
\usepackage{balance}
\usepackage[caption=false,font=footnotesize]{subfig}
\usepackage{multirow}
\usepackage{float}
\usepackage{pdfpages}
\usepackage[capitalise]{cleveref}
\usepackage{amsmath}
\usepackage{amssymb}
\newcommand{\SetCapsType}{normalcaps}

\makeatletter
\providecommand{\SetCapsType}{smallcaps}

\long\def\@scTrue{smallcaps}
\long\def\@scFalse{normalcaps}
\newcommand{\acroSCaps}[1]{%
 \begingroup
  \ifx\SetCapsType\@scTrue 
    \textsc{#1}%
  \else
    \MakeUppercase{#1}%
  \fi
  \endgroup
}
\makeatother

\newcommand{\nAcronym}[4][]{%
	\newacronym[#1]{#2}{#3}{#4}
}

\makeatletter
\@ifpackageloaded{babel}{%
    \newcommand{\usuk}[2]{%
        \iflanguage{USenglish}{#1}{#2}%
    }%
}{%
    \newcommand{\usuk}[2]{%
        #1%
    }%
}%
\makeatother

\usepackage[shortcuts]{extdash}
\newcommand{\qam}[1]{
    \ifglsused{QAM}%
        {#1\=/\gls{QAM}}%
        {#1\=/ary \gls{QAM}%
    }%
}%

\newcommand{\pam}[1]{
    \ifglsused{PAM}%
        {#1\=/\gls{PAM}}%
        {#1\=/ary \gls{PAM}%
    }%
}%

\nAcronym{2A8PSK}{\acroSCaps{2a8psk}}{2-ary amplitude 8-ary phaseshift keying}

\nAcronym{3CCMCF}{\acroSCaps{3cc-mcf}}{3-core coupled-core multi-core fiber}

\nAcronym{4D}{\acroSCaps{4d}}{four-dimensional}
\nAcronym{4D64PRS}{\acroSCaps{4d-64prs}}{\usuk{four-dimensional 64-ary polarization-ring-switching}{four-dimensional 64-ary polarisation-ring-switching}}

\nAcronym{5B4D2A8PSK}{\acroSCaps{5b4d-2a8psk}}{5-bit four-dimensional two-amplitude 8-ary phase-shift keying}

\nAcronym{6B4D2A8PSK}{\acroSCaps{6b4d-2a8psk}}{6-bit four-dimensional two-amplitude 8-ary phase-shift keying}

\nAcronym{8D}{\acroSCaps{8d}}{eight-dimensional}\nAcronym{8D2048PRS}{\acroSCaps{8d-2048prs}}{eight-dimensional 2048-ary polarization-ring-switching}
\nAcronym{8D2048PRST1}{\acroSCaps{8d-2048prs-t1}}{eight-dimensional 2048-ary polarization-ring-switching type 1}
\nAcronym{8D2048PRST2}{\acroSCaps{8d-2048prs-t2}}{eight-dimensional 2048-ary polarization-ring-switching type 2}

\nAcronym{ABC}{\acroSCaps{abc}}{automatic bias controller}
\nAcronym{ADC}{\acroSCaps{adc}}{analog-to-digital converter}
\nAcronym{AIR}{\acroSCaps{air}}{achievable information rate}
\nAcronym{AOM}{\acroSCaps{aom}}{acoustic optical modulator}
\nAcronym{AR}{\acroSCaps{ar}}{achievable rate}
\nAcronym{ASE}{\acroSCaps{ase}}{amplified spontaneous emission}
\nAcronym{ASIC}{\acroSCaps{asic}}{application-specific integrated circuit}
\nAcronym{AWGN}{\acroSCaps{awgn}}{additive white Gaussian noise}

\nAcronym{BER}{\acroSCaps{ber}}{bit error rate}
\nAcronym{BICM}{\acroSCaps{bicm}}{bit-interleaved coded modulation}
\nAcronym{BMD}{\acroSCaps{bmd}}{bit-metric decoding}
\nAcronym{BPD}{\acroSCaps{bpd}}{balanced photo-diode}
\nAcronym{BPS}{\acroSCaps{bps}}{blind phase search}
\nAcronym{BRGC}{\acroSCaps{brgc}}{binary reflected Gray code}
\nAcronym{BTB}{\acroSCaps{btb}}{back-to-back}

\nAcronym{CCDM}{\acroSCaps{ccdm}}{constant composition distribution matching}
\nAcronym{CCF}{\acroSCaps{ccf}}{\usuk{coupled-core fiber}{coupled-core fibre}}
\nAcronym{CD}{\acroSCaps{cd}}{chromatic dispersion}
\nAcronym{CIR}{\acroSCaps{cir}}{channel impulse response}
\nAcronym{CMUX}{\acroSCaps{cmux}}{core multiplexer}
\nAcronym{ChUT}{\acroSCaps{chut}}{channel under test}
\nAcronym{CUT}{\acroSCaps{cut}}{channel under test}
\nAcronym{CPE}{\acroSCaps{cpe}}{carrier phase estimation}
\nAcronym{CAGR}{\acroSCaps{cagr}}{compound annual growth rate}

\nAcronym{DA}{\acroSCaps{da}}{driver amplifier}
\nAcronym{DAC}{\acroSCaps{dac}}{digital-to-analog converter}
\nAcronym{DCF}{\acroSCaps{dcf}}{\usuk{dispersion compensated fiber}{dispersion compensated fibre}}
\nAcronym{DCI}{\acroSCaps{dci}}{data center interconnect}
\nAcronym{DFB}{\acroSCaps{dfb}}{distributed feedback}
\nAcronym{DGD}{\acroSCaps{dgd}}{differential group delay}
\nAcronym{DM}{\acroSCaps{dm}}{distribution matcher}
\nAcronym{DMGD}{\acroSCaps{dmgd}}{differential mode group delay}
\nAcronym{DP}{\acroSCaps{dp}}{\usuk{dual-polarization}{dual-polarisation}}
\nAcronym{DPC}{\acroSCaps{dpc}}{digital pre-compensation}
\nAcronym{DPE}{\acroSCaps{dpe}}{digital pre-emphasis}
\nAcronym{DPIQ}{\acroSCaps{dp-iqm}}{\usuk{dual-polarization IQ-modulator}{dual-polarisation IQ-modulator}}
\nAcronym{DRE}{\acroSCaps{dre}}{digital resolution enhancer}
\nAcronym{DSO}{\acroSCaps{dso}}{digital sampling oscilloscope}
\nAcronym{DUT}{\acroSCaps{dut}}{device-under-test}
\nAcronym{DWDM}{\acroSCaps{dwdm}}{dense wavelength-division multiplexing}

\nAcronym{ED}{\acroSCaps{ed}}{Eucledian distance}
\nAcronym{ENOB}{\acroSCaps{enob}}{effective number of bits}
\nAcronym{ESS}{\acroSCaps{ess}}{enumerative sphere shaping}

\nAcronym{FDE}{\acroSCaps{fde}}{\usuk{frequency domain equalizer}{frequency domain equaliser}}
\nAcronym{FEC}{\acroSCaps{fec}}{forward error correction}
\nAcronym{FFT}{\acroSCaps{fft}}{fast Fourier transform}
\nAcronym{FWM}{\acroSCaps{fwm}}{four-wave mixing}

\nAcronym{GFF}{\acroSCaps{gff}}{gain flattening filter}
\nAcronym{GIMMF}{\acroSCaps{gi-mmf}}{\usuk{graded-index multi-mode fiber}{graded-index multi-mode fibre}}
\nAcronym{GMI}{\acroSCaps{gmi}}{\usuk{generalized mutual information}{generalised mutual information}}
\nAcronym{GS}{\acroSCaps{gs}}{geometric shaping}
\nAcronym{GV}{\acroSCaps{gv}}{group-velocity}
\nAcronym{GVD}{\acroSCaps{gvd}}{group-velocity dispersion}

\nAcronym{HDFEC}{\acroSCaps{hd-fec}}{hard decision forward error correction}

\nAcronym[plural=IL, firstplural=insertion losses (\textsc{il})]{IL}{\acroSCaps{il}}{insertion loss}
\nAcronym{IFFT}{\acroSCaps{ifft}}{inverse fast Fourier transform}
\nAcronym{ISI}{\acroSCaps{isi}}{intersymbol interference}

\nAcronym{KK}{\acroSCaps{kk}}{Kramers-Kronig}

\nAcronym{LCOS}{\acroSCaps{LCoS}}{liquid crystal on silicon}
\nAcronym{LDPC}{\acroSCaps{ldpc}}{low-density parity-check}
\nAcronym{LEAF}{\acroSCaps{leaf}}{\usuk{large effective area fiber}{large effective area fibre}}
\nAcronym{LLR}{\acroSCaps{llr}}{log-likelihood ratio}
\nAcronym{LO}{\acroSCaps{lo}}{local oscillator}
\nAcronym{LSPS}{\acroSCaps{lsps}}{\usuk{loop-synchronized polarization scrambler}{loop-synchronised polarisation scrambler}}
\nAcronym{LUT}{\acroSCaps{lut}}{lookup table}

\nAcronym{MB}{\acroSCaps{mb}}{Maxwell-Bolzmann}
\nAcronym{MDM}{\acroSCaps{mdm}}{mode division multiplexing}
\nAcronym{MF}{\acroSCaps{mf}}{matched filter}
\nAcronym{MI}{\acroSCaps{mi}}{mutual information}
\nAcronym{MMA}{\acroSCaps{mma}}{multi-modulus algorithm}
\nAcronym{MPLC}{\acroSCaps{mplc}}{multi-plane light converter}
\nAcronym{MSE}{\acroSCaps{mse}}{mean squared error}
\nAcronym{MUX}{\acroSCaps{mux}}{multiplexer}

\nAcronym{NF}{\acroSCaps{nf}}{noise figure}
\nAcronym{NGMI}{\acroSCaps{ngmi}}{\usuk{normalized generalized mutual information}{normalised generalised mutual information}}
\nAcronym{NIR}{\acroSCaps{nir}}{near infra-red}

\nAcronym{OBTB}{\acroSCaps{obtb}}{optical back-to-back}
\nAcronym{ODL}{\acroSCaps{odl}}{optical delay line}
\nAcronym{OFDR}{\acroSCaps{ofdr}}{optical frequency-domain reflectometer}
\nAcronym{OFDM}{\acroSCaps{ofdm}}{orthogonal frequency division multiplexing}
\nAcronym{OMFT}{\acroSCaps{omft}}{optical-multi-format transmitter}
\nAcronym{OSA}{\acroSCaps{osa}}{\usuk{optical spectrum analyzer}{optical spectrum analyser}}
\nAcronym{OTDR}{\acroSCaps{otdr}}{optical time-domain reflectometer}
\nAcronym{OTF}{\acroSCaps{otf}}{optical tunable filter}
\nAcronym{OVNA}{\acroSCaps{ovna}}{\usuk{optical vector network analyzer}{optical vector network analyser}}

\nAcronym{PAM}{\acroSCaps{pam}}{pulse-amplitude modulation}
\nAcronym{PAS}{\acroSCaps{pas}}{probabilistic amplitude shaping}
\nAcronym{PAPR}{\acroSCaps{papr}}{peak-to-avarage power ratio}
\nAcronym{PBS}{\acroSCaps{pbs}}{polarization beam splitter}
\nAcronym{PCVD}{\acroSCaps{pcvd}}{plasma chemical vapor depostion}
\nAcronym{PDL}{\acroSCaps{pdl}}{polarization-dependent loss}
\nAcronym{PDM}{\acroSCaps{pdm}}{\usuk{polarization-division multiplexing}{polarisation-division multiplexing}}
\nAcronym{PMBPSK}{\acroSCaps{pm-bpsk}}{polarization-multiplexed binary phase-shift-keying}
\nAcronym{PMQPSK}{\acroSCaps{pm-qpsk}}{polarization-multiplexed quaternary phase-shift-keying}
\nAcronym{PM8QAM}{\acroSCaps{pm-8qam}}{polarization-multiplexed 8-ary quadrature amplitude modulation}
\nAcronym{PMD}{\acroSCaps{pmd}}{polarization mode dispersion}
\nAcronym{PNOB}{\acroSCaps{pnob}}{physical number of bits}
\nAcronym{PRBS}{\acroSCaps{prbs}}{pseudorandom bit sequence}
\nAcronym{PS}{\acroSCaps{ps}}{probabilistic shaping}

\nAcronym{RF}{\acroSCaps{rf}}{radio frequency}
\nAcronym{SamPerSym}{\acroSCaps{sps}}{samples per symbol}
\nAcronym{SDFEC}{\acroSCaps{sd-fec}}{soft decision forward error correction}
\nAcronym{SE}{\acroSCaps{se}}{spectral efficiency}
\nAcronym{SER}{\acroSCaps{ser}}{symbol error rate}
\nAcronym{SA}{\acroSCaps{sa}}{simulated annealing}
\nAcronym{SLM}{\acroSCaps{slm}}{spatial light modulator}
\nAcronym{SSFM}{\acroSCaps{ssfm}}{split-step Fourier method}
\nAcronym{SMF}{\acroSCaps{smf}}{\usuk{single-mode fiber}{single-mode fibre}}
\nAcronym[firstplural=\usuk{states of polarization (\acroSCaps{sop})}{states of polarisation (\acroSCaps{sop})}]{SOP}{\acroSCaps{sop}}{\usuk{state of polarization}{state of polarisation}}
\nAcronym{SPM}{\acroSCaps{spm}}{self-phase modulation}
\nAcronym{SPS}{\acroSCaps{sps}}{\usuk{synchronized polarization scrambler}{synchronised polarisation scrambler}}
\nAcronym{SSMF}{\acroSCaps{ssmf}}{\usuk{standard single-mode fiber}{standard single-mode fibre}}
\nAcronym{SVD}{\acroSCaps{svd}}{singular value decomposition}
\nAcronym{SWI}{\acroSCaps{swi}}{swept wavelength interferometry}

\nAcronym{TDE}{\acroSCaps{tde}}{\usuk{time domain equalizer}{time domain equaliser}}
\nAcronym{TDMSDM}{\acroSCaps{tdm-sdm}}{time-domain multiplexed space-division multiplexing}
\nAcronym{TE}{\acroSCaps{TE}}{transverse electric}
\nAcronym{TM}{\acroSCaps{TM}}{transverse magnetic}
\nAcronym{TH4D}{\acroSCaps{th-4d}}{time domain hybrid four-dimensional}
\nAcronym{TH4D2A8PSK}{\acroSCaps{th-4d-2a8psk}}{time domain hybrid four-dimensional two-amplitude eight-phase shift keying}

\nAcronym{VOA}{\acroSCaps{voa}}{variable optical attenuator}

\nAcronym{WSS}{\acroSCaps{wss}}{wavelength selective switch}

\nAcronym{XPM}{\acroSCaps{xpm}}{cross-phase modulation}
\nAcronym{XT}{\acroSCaps{xt}}{cross-talk}

\nAcronym{3DWG}{\acroSCaps{3dwg}}{3D-waveguide}

\nAcronym{DD}{\acroSCaps{dd}}{decision-directed}
\nAcronym{DMD}{\acroSCaps{dmd}}{differential-mode delay}
\nAcronym{DSP}{\acroSCaps{dsp}}{digital signal processing}

\nAcronym{ECL}{\acroSCaps{ecl}}{external cavity laser}
\nAcronym{EDFA}{\acroSCaps{edfa}}{\usuk{erbium-doped fiber amplifier}{erbium-doped fibre amplifier}}

\nAcronym{GD}{\acroSCaps{gd}}{group delay}
\nAcronym{GIFMF}{\acroSCaps{gi-fmf}}{\usuk{graded-index few-mode fiber}{graded-index few-mode fibre}}

\nAcronym{LP}{\acroSCaps{lp}}{\usuk{linearly polarized}{linearly polarised}}
\nAcronym{LMS}{\acroSCaps{lms}}{least mean square}
\nAcronym{MMSE}{\acroSCaps{mmse}}{minimum mean square error}
\nAcronym{MCF}{\acroSCaps{mcf}}{\usuk{multi-core fiber}{multi-core fibre}}
\nAcronym{MDG}{\acroSCaps{mdg}}{mode-dependent gain}
\nAcronym{MDL}{\acroSCaps{mdl}}{mode-dependent loss}
\nAcronym{MMF}{\acroSCaps{mmf}}{\usuk{multi-mode fiber}{multi-mode fibre}}
\nAcronym{MZM}{\acroSCaps{mzm}}{Mach-Zehnder modulator}
\nAcronym{MIMO}{\acroSCaps{mimo}}{multiple-input multiple-output}

\nAcronym{FMF}{\acroSCaps{fmf}}{\usuk{few-mode fiber}{few-mode fibre}}
\nAcronym{FM-MCF}{\acroSCaps{fm-mcf}}{\usuk{few-mode multi-core fiber}{few-mode multi-core fiber}}

\nAcronym{OSNR}{\acroSCaps{osnr}}{optical signal-to-noise ratio}

\nAcronym{PL}{\acroSCaps{pl}}{photonic lantern}

\nAcronym{RRC}{\acroSCaps{rrc}}{root-raised-cosine}

\newacronym{SDM}{SDM}{space division multiplexing}
\nAcronym{SNR}{\acroSCaps{snr}}{signal-to-noise ratio}

\nAcronym{ULH}{\acroSCaps{ulh}}{ultra-long-haul}

\nAcronym{QAM}{\acroSCaps{qam}}{quadrature amplitude modulation}
\nAcronym{QPSK}{\acroSCaps{qpsk}}{quaternary phase-shift-keying}

\nAcronym{WDM}{\acroSCaps{wdm}}{wavelength-division multiplexing}

\nAcronym{CombIR}{\acroSCaps{cir}}{combined impulse response}

\usepackage[]{xcolor}

\newcommand{\LP}[1]{LP\textsubscript{#1}}

\usepackage{silence}
\begin{document}
\bstctlcite{IEEEexampleLimitAuthors:BSTcontrol}
%


    \title{Mode-dependent Loss and Gain Estimation in SDM Transmission Based on MMSE Equalizers}

%
%
%

    \author{Ruby~S.~B.~Ospina,~\IEEEmembership{Student Member,~OSA,}
        Menno~van~den~Hout,~\IEEEmembership{Student Member,~IEEE,}\\
        Juan~Carlos~Alvarado-Zacarias,~\IEEEmembership{Student Member,~IEEE,}
        Jose~Enrique~Antonio-L\'opez,
        Marianne~Bigot-Astruc,
        Adrian~Amezcua~Correa,
        Pierre~Sillard,~\IEEEmembership{Member,~IEEE},
        Rodrigo~Amezcua-Correa,~\IEEEmembership{Member,~IEEE,}\\
        ~Chigo~Okonkwo,~\IEEEmembership{Senior Member,~IEEE,}
        and Darli~A.~A.~Mello,~\IEEEmembership{Member,~IEEE}%
        \vspace{-4mm}
        \thanks{Manuscript received XXX xx, XXXX; revised XXXXX xx, XXXX; accepted XXXX XX, XXXX. Date of publication XXXX XX, XXXX. This work was partially supported by FAPESP under grants 2018/25414-6, 2018/14026-5, 2017/25537-8, 2015/24341-7, 2015/ 24517-8 and by the TU/e-KPN flagship Smart Two project. This article was presented in part at the Optical Fiber Communications Conference, San Diego, CA, USA, Mar. 2020.}
        \thanks{R. S. B. Ospina and D. A. A. Mello are with the School of Electrical and Computer Engineering, State University of Campinas, Campinas 13083-970, Brazil (e-mail: r163653@dac.unicamp.br, darli@unicamp.br).}%
        \thanks{M. van den Hout and C. Okonkwo are with the High Capacity Optical Transmission Laboratory, Electro-Optical Communications Group, Eindhoven University of Technology, PO Box 513, 5600 MB, Eindhoven, The Netherlands. (e-mail: \{m.v.d.hout; c.m.okonwko\}@tue.nl).}%
        \thanks{J.C. Alvarado-Zacarias, J.E. Antonio-L\'opez and R. Amezcua-Correa are with the CREOL, The College of Optics and Photonics, University of Central Florida, Orlando, 32816, USA (e-mail: {jcalvarazac@knights.ucf.edu, jealopez@creol.ucf.edu, r.amezcua@creol.ucf.edu})}
        \thanks{M. Bigot-Astruc, A. Amezcua Correa and P. Sillard are with Prysmian Group, 644 Boulevard Est, Billy Berclau, 62092 Haisnes Cedex, France. (e-mail: \{Marianne.Bigot; Adrian.amezcua; Pierre.Sillard\}@prysmiangroup.com)}
        \thanks{Color versions of one or more figures in this paper are available online at http://ieeexplore.ieee.org.}
        \thanks{Digital Object Identifier xxxxxxxxxxxxxxxxxxxxxxxx}
    }

    \IEEEpubid{}

    \IEEEpubid{\begin{minipage}{\textwidth}
                   \ \\[12pt]
                   \centering
                   0000--0000/00\$00.00~\copyright~2020 IEEE Personal use is permitted, but republication/redistribution requires IEEE permission.\\See http://www.ieee.org/publications standards/publications/rights/index.html for more information.
    \end{minipage}}



    \maketitle

    \begin{abstract}
        The capacity in \gls{SDM} systems with coupled channels is fundamentally limited by \gls{MDL} and \gls{MDG} generated in components and amplifiers. In these systems, \gls{MDL}/\gls{MDG} must be accurately estimated for performance analysis and troubleshooting. Most recent demonstrations of \gls{SDM} with coupled channels perform \gls{MDL}/\gls{MDG} estimation by \gls{DSP} techniques based on the coefficients of \gls{MIMO} adaptive equalizers. Although these methods provide a valid indication of the order of magnitude of the accumulated MDL/MDG over the link, \gls{MIMO} equalizers are usually updated according to the \gls{MMSE} criterion, which is known to depend on the channel \gls{SNR}. Therefore, \gls{MDL}/\gls{MDG} estimation techniques based on the adaptive filter coefficients are also impaired by noise. In this paper, we model analytically the influence of the \gls{SNR} on DSP-based \gls{MDL}/\gls{MDG} estimation, and show that the technique is prone to errors. Based on the transfer function of MIMO MMSE equalizers, and assuming a known \gls{SNR}, we calculate a correction factor that improves the estimation process in moderate levels of \gls{MDL}/\gls{MDG} and \gls{SNR}. The correction factor is validated by simulation of a 6-mode long-haul transmission link, and experimentally using a 3-mode transmission link. The results confirm the limitations of the standard estimation method in scenarios of high additive noise and \gls{MDL}/\gls{MDG}, and indicate the correction factor as a possible solution in practical \gls{SDM} scenarios.
    \end{abstract}

    \begin{IEEEkeywords}
        Mode-dependent loss, mode-dependent gain, space division multiplexing, optical fiber communications.
    \end{IEEEkeywords}

    \glsreset{SDM}
    \glsreset{MDL}
    \glsreset{MDG}
    \glsreset{SNR}
    \glsreset{MIMO}
    \glsreset{MMSE}
    \glsreset{DSP}

%
    \IEEEpeerreviewmaketitle

    \section{Introduction}
    \IEEEPARstart{S}{pace} division multiplexing (\acrshort{SDM}) enables significant increase in capacity and integration at the component, fiber and system level \cite{winzer2014optical}. In recent years, several \acrshort{SDM} flavors have been proposed over single-mode fiber bundles \cite{AlaelsonJatoba2018}, \cite{OmarDomingues2017}, uncoupled or coupled \glspl{MCF} \cite{igarashi2015114}, \cite{ryf2019coupled}, \glspl{MMF}, \glspl{FMF} \cite{van2018138}, or \glspl{FM-MCF} \cite{soma201710}.
    SDM long-haul transmission with coupled channels demands \gls{MIMO} equalizers at reception to compensate for any linear mixing between modes. In addition to linear coupling and modal dispersion, the guided modes may be subject to unequal attenuation and amplification. This effect is known as \gls{MDL} and \gls{MDG}.
    \IEEEpubidadjcol
    \gls{MDL}/\gls{MDG} turn the channel capacity into a random variable, reducing the average channel capacity and generating outages \cite{ho2011mode}. The combined effect of accumulated MDL/MDG and amplifier noise fundamentally limits the performance of high-capacity SDM systems to be deployed at long distances. 
    The impact of MDL/MDG on the channel capacity of coupled SDM transmission has been widely investigated. In \cite{ho2011mode}, Ho \textit{et al.} present a statistical characterization of MDL and quantify its effect on the channel capacity in strongly coupled SDM systems. In \cite{winzer2011mimo}, Winzer \textit{et al.} discuss the MDL-induced capacity reduction in SDM channels and provide closed-form expressions for the system outage performance. In \cite{8918466}, Mello \textit{et al.} review analytical expressions for channel capacity in \gls{MDG}-impaired SDM systems. Moreover, they study the effect of frequency diversity on the \gls{MDG}-induced outage probability and quantify the maximum tolerable per-amplifier \gls{MDG} for a certain average capacity metric in \gls{ULH} systems.

\begin{figure*}[t]
        \centering
        \includegraphics[width=0.92\linewidth]{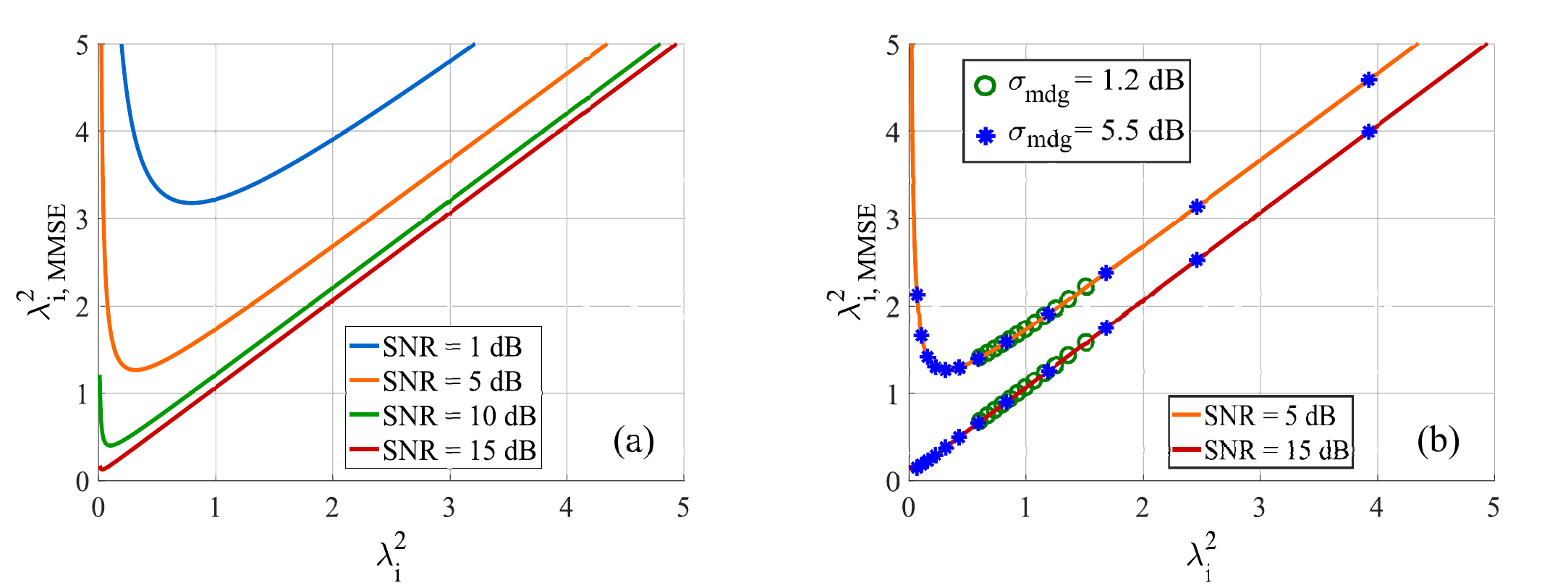}%
        \caption{(a) Eigenvalues estimated by \gls{DSP}, $\lambda^{2}_{i_{\mathrm{MMSE}}}$, as a function of the actual eigenvalues $\lambda^{2}_{i}$ for different levels of \gls{SNR}. (b) Distribution of the 12 eigenvalues of a 6 spatial modes channel at two different levels of MDL/MDG and SNR.}%
        \label{fig:lambdarelation}
    \end{figure*}

    \Acrfull{DSP}-based estimation of \gls{MDL}/\gls{MDG} by coherent receivers yields a two-fold benefit namely: assessing the link performance and estimating a lower bound on the per-amplifier \gls{MDG} performance. In DSP-based \gls{MDL}/\gls{MDG} estimation, the channel transfer function is usually estimated by the inverse frequency response of the MIMO equalizer. In \cite{van2018138}, Van Weerdenburg \textit{et al.} presents the evolution with distance of the \gls{MDL} estimated with a 12 $\times$ 12 \gls{LMS} MIMO equalizer in a 138-Tb/s 6-mode transmission. In \cite{van2017138}, the same authors discuss the DSP-based MDL estimation over 120 wavelength channels throughout the C-band in a 650 km 6-mode transmission. In \cite{rademacher2018long}, Rademacher \textit{et al.} use the coefficients of a 6~$\times$~6 MIMO equalizer to estimate the MDL at different modal launch powers in a 3-mode transmission. Also, recently in \cite{rademacher202010}, Rademacher \textit{et al.} estimate the \gls{MDL} in a 38-core-3-mode transmission by employing a 6 $\times$ 6 MIMO equalizer to process each core.
    
    We showed in \cite{ospina2020dsp} that, as adaptive \gls{MIMO} equalizers typically use the \gls{MMSE} criterion \cite{faruk2017digital}, the \gls{MDL}/\gls{MDG} estimation accuracy is affected by the channel \acrfull{SNR}.
    Using the analytical transfer function of MMSE equalizers, we also show that \gls{MDL}/\gls{MDG} estimation errors can be partially compensated by a correction factor based on a known SNR. The results are validated by Monte-Carlo simulation. In this paper, we extend the results of \cite{ospina2020dsp}, by addressing the problem in more detail and validating the method experimentally using a 3-mode transmission link.

    The remainder of this paper is divided as follows. Section II covers the fundamentals of DSP-based \gls{MDL}/\gls{MDG} estimation and derives the correction factor adopted to enhance the estimation accuracy. Section III presents the validation of the correction factor by simulations of a 6-mode long-haul transmission. Section IV describes the experimental setup for a 3-mode transmission link. Section V presents the experimental results obtained with the 3-mode transmission link. Lastly, Section VI concludes the paper.

    \section{DSP-based mode-dependent loss and gain estimation}
    \label{sec:dsp estimation}

    \acrshort{SDM} optical systems with coupled channels deployed at long distances will require multiple inline amplifiers that may present a different gain profile for the various guided spatial modes, resulting in \gls{MDG}. Moreover, fiber bends, connectors, splices, and optical devices, such as multiplexers, also introduce \gls{MDL}.
    The \gls{MDL}/\gls{MDG} of a link can be computed from the eigenvalues $\lambda_{i}^{2}$ of the operator $\mathbf{H}\mathbf{H}^{H}$, where $\mathbf{H}$ is the channel transfer matrix, and $(.)^{H}$ denotes the Hermitian transpose operator \cite{winzer2011mimo,ho2011mode}. \gls{MDL}/\gls{MDG} is usually quantified by two possible metrics. In links with weak mode coupling, the peak-to-peak value, given by the ratio between the highest and the lowest eigenvalues in dB, is a relevant metric \cite{winzer2011mimo}, \cite{lobato2013mode}. On the other hand, the standard deviation of the eigenvalues in logarithmic scale, $\sigma_{\mathrm{mdg}}$, is usually employed to characterize the \gls{MDL}/\gls{MDG} in links with strong mode coupling \cite{ho2011mode}, \cite{8918466}, \cite{choutagunta2018characterizing}. In this paper, we focus on the standard deviation metric because of its direct applicability in long-distance links \cite{choutagunta2018characterizing}.
    In DSP-based \gls{MDL}/\gls{MDG} estimation, the channel transfer matrix $\mathbf{H}$ is usually estimated from the MIMO equalizer transfer function $\mathbf{W}$ \cite{van2018138}, \cite{rademacher2018long}, \cite{rademacher202010}. The MMSE equalizer transfer function can be expressed as \cite{kim2008performance}, \cite{mckay2009achievable}
    \begin{equation}
        \mathbf{W}_{\mathrm{MMSE}} = \left( \frac{\mathrm{\mathbf{I}}}{\mathrm{SNR}} + \mathbf{H}^{H}\mathbf{H} \right)^{-1}\mathbf{H}^{H}.
        \label{Eq: wmmse}
    \end{equation}
    \noindent The SNR in (\ref{Eq: wmmse}) is the electrical SNR calculated before the MIMO equalizer, and is equivalent to the optical SNR (OSNR) calculated using the signal bandwidth as reference bandwidth. As \gls{DSP}-based \gls{MDL}/\gls{MDG} estimation uses $\mathbf{W}^{-1}_{\mathrm{MMSE}}$ as an estimate of $\mathbf{H}$, the eigenvalues $\lambda_{i}^{2}$ are estimated from the eigenvalues of $\mathbf{W}^{-1}_{\mathrm{MMSE}}(\mathbf{W}^{-1}_{\mathrm{MMSE}})^H$. The relationship between the actual eigenvalues $\lambda_{i}^{2}$ and the eigenvalues obtained by \gls{DSP}, $\lambda^{2}_{i_{\mathrm{MMSE}}}$, can be obtained from the eigendecomposition of  $\mathbf{W}^{-1}_{\mathrm{MMSE}}(\mathbf{W}^{-1}_{\mathrm{MMSE}})^H$, as
    \begin{equation}
        \begin{split}
            \mathbf{W}^{-1}_{\mathrm{MMSE}}(\mathbf{W}^{-1}_{\mathrm{MMSE}})^H &= \frac{\left(\mathbf{H}\mathbf{H}^{H} \right)^{-1}}{\mathrm{SNR}^{2}}+\frac{2\mathbf{I}}{\mathrm{SNR}}+\mathbf{H}\mathbf{H}^{H} \\
            & = \mathbf{Q}  \left[ \frac{\mathbf{\Lambda_{\mathbf{\mathrm{H}}}}^{-1}}{\mathrm{SNR}^{2}} + \frac{2\mathbf{I}}{\mathrm{SNR}} + \mathbf{\Lambda_{\mathbf{\mathrm{H}}}} \right]   \mathbf{Q^{-1}},
            \label{Eq:eigendes1}
        \end{split}
    \end{equation}
    \noindent where $\mathbf{\Lambda_{\mathbf{\mathrm{H}}}}$ is a diagonal matrix whose main diagonal has elements $\lambda_{i}^{2}$, and $\mathbf{Q}$ is a matrix whose columns are the eigenvectors of $\mathbf{H}\mathbf{H}^H$. From (\ref{Eq:eigendes1}), the original eigenvalues $\lambda_{i}^{2}$, and the eigenvalues obtained by DSP, $\lambda^{2}_{i_{\mathrm{MMSE}}}$, are related by \cite{ospina2020dsp}
    \begin{equation}
        \lambda^{2}_{i_{\mathrm{MMSE}}} = \left[\frac{\left(\lambda^{2}_{i}\right)^{-1}}{\mathrm{SNR}^{2}} + \frac{2}{\mathrm{SNR}} + \lambda^{2}_{i} \right],
        \label{Eq:Lambdarelation}
    \end{equation}

    \noindent \cref{fig:lambdarelation}a shows $\lambda^{2}_{i_{\mathrm{MMSE}}}$ as a function of $\lambda_{i}^{2}$ for different levels of \gls{SNR}.
    At high \gls{SNR}, the first term in (\ref{Eq:Lambdarelation}) tends to zero, and $\lambda_{i}^{2}$ and $\lambda^{2}_{i_{\mathrm{MMSE}}}$ are linearly related with linear coefficient $\mathrm{2/}$SNR. As the \gls{SNR} decreases, lower values of $\lambda_{i}^{2}$ start to raise, breaking the linear relation, further impairing the estimation process. \cref{fig:lambdarelation}b illustrates the same effect, indicating by markers the eigenvalues obtained by different realizations of \gls{MDL}/\gls{MDG} and SNR in a system with 6 spatial modes (12 spatial and polarization modes). At SNR = 15 dB, the eigenvalues are positioned on the $x=y$ curve, and the conventional estimation process is successful. At SNR = 5 dB and $\sigma_{\mathrm{mdg}} = 1.2$ dB, the estimated eigenvalues are simply displaced by $2/$SNR. If the SNR is known, this displacement can be corrected. At SNR = 5 dB and $\sigma_{\mathrm{mdg}} = 5.5$ dB, the eigenvalues $\lambda_{i}^{2}$ in blue asterisks disperse and the lower estimated eigenvalues raise because of the nonlinear term. In this condition, the estimation accuracy of the standard DSP-based method is strongly affected.

    If the \gls{SNR} is known, (\ref{Eq:Lambdarelation}) can be inverted to recover $\lambda_{i}^{2}$ from $\lambda^{2}_{i_{\mathrm{MMSE}}}$, resulting in a quadratic equation with two roots
    \begin{equation}
        \label{Eq:roots}
        \resizebox{1\hsize}{!}{$%
        \lambda^{2}_{i}= \frac{ \left[ \mathrm{SNR}^{2}\,\lambda^{2}_{i_{\mathrm{MMSE}}}-2\,\mathrm{SNR} \right] {\pm} \sqrt{\left[\mathrm{SNR}^{2}\,\lambda^{2}_{i_{\mathrm{MMSE}}}-2\,\mathrm{SNR} \right]^2 - 4\,\mathrm{SNR}^{2}} }{2\,\mathrm{SNR}^{2}}.
        $}
    \end{equation}

    The highest solution of (\ref{Eq:roots}) recovers $\lambda^{2}_{i}$ for high and moderate values of \gls{SNR} and low and moderate values of \gls{MDL}/\gls{MDG}. In this paper, we adopt such positive solution as a correction factor applied over the DSP eigenvalues, $\lambda^{2}_{i_{\mathrm{MMSE}}}$, to enhance the \gls{MDL}/\gls{MDG} estimation process.

   The expression for the MMSE equalizer in (\ref{Eq: wmmse}) and the correction of the estimated eigenvalues by (\ref{Eq:roots}) apply to any coupled system represented by a transfer matrix $\mathbf{H}$, irrespective if it operates in the regimes of weak or strong mode coupling. In this paper, we evaluate a long-haul transmission system with strong-mode coupling through simulations, and a short-reach transmission system with weak mode coupling through experiments. 

    \section{Simulation results for long-haul transmission }
\begin{figure}[htb]
        \captionsetup[subfigure]{labelformat=empty}
        \centering
        \subfloat[\label{subfig:contoura}]{%
            \includegraphics[width=0.99\linewidth]{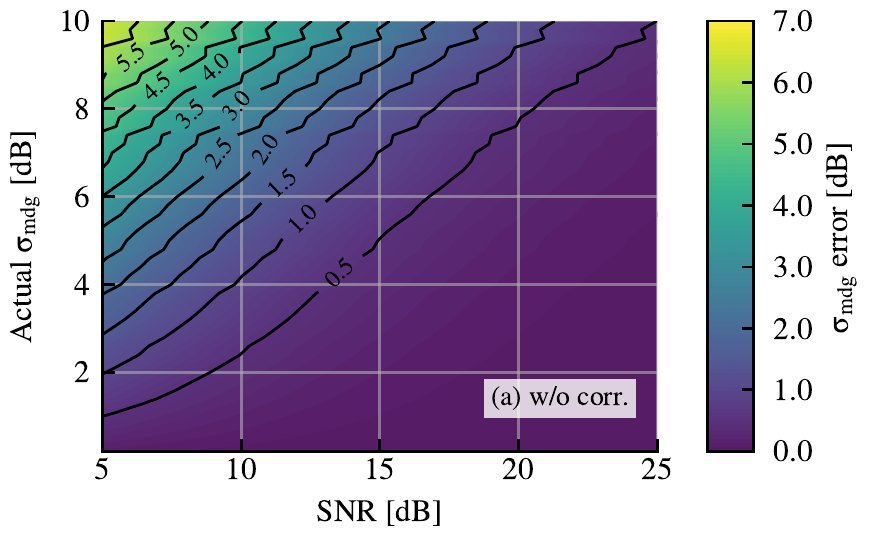}}
        \\
        \subfloat[\label{subfig:contourb}]{%
            \includegraphics[width=0.99\linewidth]{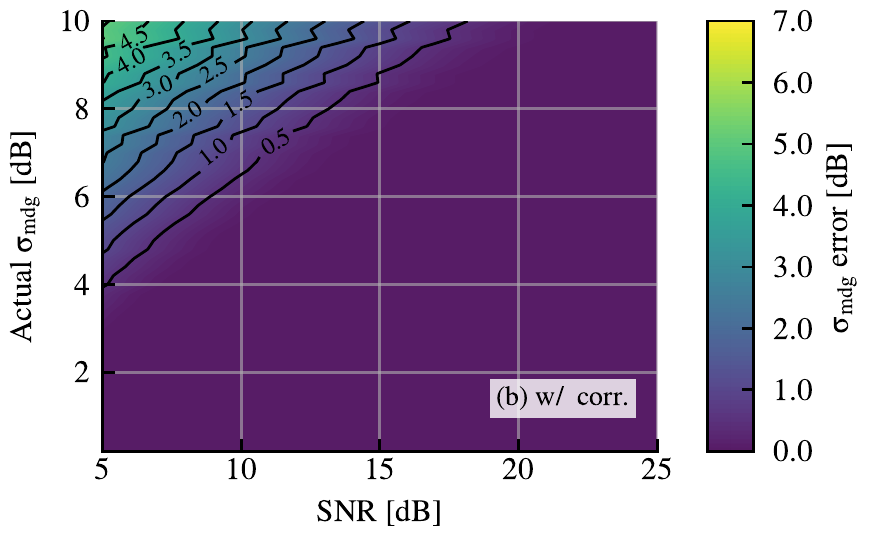}}
        \vspace*{-3mm}
        \caption{Analytical estimation error in dB computed as the absolute value of the difference between the actual MDL/MDG standard deviation $\sigma_{\mathrm{mdg}}$, and the $\sigma_{\mathrm{mdg}}$ estimated in dB using $\mathbf{W}^{-1}_{\mathrm{MMSE}}$ computed from $\mathbf{H}$ using (\ref{Eq: wmmse}). (a) Without correction. (b) Correction by the positive solution of (\ref{Eq:roots}).}
        \label{fig:contour}
    \end{figure}

    Firstly, we analytically evaluate the performance of \gls{DSP}-based \gls{MDL}/\gls{MDG} estimation in long-haul SDM transmission. Using the multisection model presented in \cite{ho2011mode}, $12\times12$ matrices $\mathbf{H}$ are generated to simulate a strongly coupled pol-mux 6-mode transmission of 100 50-km spans, yielding a total length of 5,000 km. The group delay standard deviation is set to 3.1 ps/$\sqrt{\mathrm{km}}$ , which is a low value for coupled core MCFs \cite{ryf2019coupled}, \cite{hayashi2017record}. The \gls{MDL}/\gls{MDG} of the link is controlled by a per-amplifier MDG standard deviation, $\sigma_{g}$.
    Matrices $\mathbf{H}$ are represented by 1,000 frequency bins over a bandwidth of 240 GHz to capture the effect of frequency diversity \cite{ho2011frequency}. The MDG standard deviation $\sigma_{\mathrm{mdg}}$ is estimated in dB using $\mathbf{W}^{-1}_{\mathrm{MMSE}}$ computed from $\mathbf{H}$ using (\ref{Eq: wmmse}). The total \gls{MDL}/\gls{MDG} is calculated by averaging over the 1,000 frequency bins. Figs. \ref{subfig:contoura} and \ref{subfig:contourb} show the contour plots of the estimation error in dB for a wide range of SNRs and MDG standard deviations $\sigma_{\mathrm{mdg}}$, without and with correction of the DSP-estimated eigenvalues, respectively. The estimation error is computed as the absolute difference between the actual and estimated $\sigma_{\mathrm{mdg}}$ in dB.
    In \cref{subfig:contoura}, without correction of the DSP-estimated eigenvalues, the estimation error achieves up to 6 dB for $\sigma_{\mathrm{mdg}} > 9$ dB across the low SNR region. At an SNR = 10 dB, an error higher than 1 dB is observed for $\sigma_{\mathrm{mdg}} > 4$  dB. Even at a higher SNR = 15 dB, the estimation error exceeds 1.5 dB for $\sigma_{\mathrm{mdg}} > 8$ dB. The contour plot in \cref{subfig:contoura} makes evident the SNR impact on the estimation accuracy. In \cref{subfig:contourb}, the correction of the DSP-estimated eigenvalues by the positive solution of (\ref{Eq:roots}) enhances the estimation process. Here, in the low SNR regime, the maximum error is 4.5 dB for $\sigma_{\mathrm{mdg}} > 9$ dB. At an SNR = 10 dB, an error higher than 1 dB is achieved only for $\sigma_{\mathrm{mdg}} > 7$  dB. For SNR $\geq$ 19 dB, the correction factor provides an estimation error below 0.5 dB over the entire range of evaluated $\sigma_{\mathrm{mdg}}$. 

\begin{figure*}[t!]
        \centering
        \includegraphics[width=0.87\linewidth]{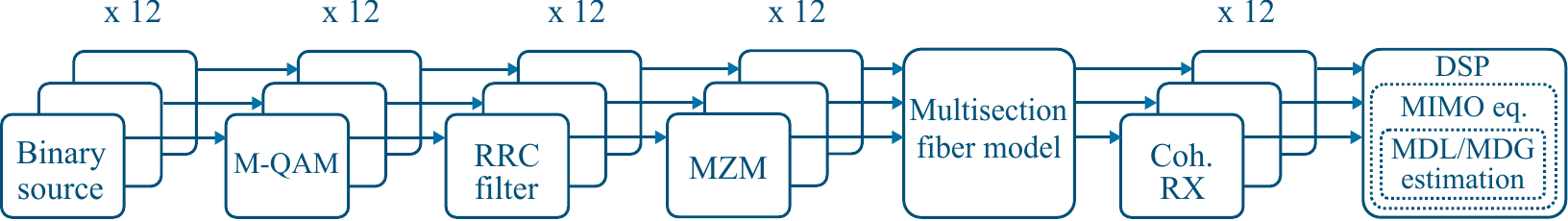}
        \caption{Simulation setup of coupled long-haul 6-mode transmission. The transmitter generates 16-QAM symbols at 30 GBd. The multisection model simulates a 5,000-km fiber link. \gls{MDL}/\gls{MDG}  estimation is performed by \gls{DSP} based on the MIMO MMSE transfer function.}
        \label{fig:simulationsetup}
    \end{figure*}

    We further simulate a coupled long-haul transmission link with $\mathrm{N_{m}}= 6$ spatial modes and polarization multiplexing, as depicted in the simulation setup of \cref{fig:simulationsetup}.
    At the transmitter, $2\,\mathrm{N_{m}}$ independent binary sequences are mapped into 400,000 16-QAM symbols at 30 GBd. The complex constellations are fed into \gls{RRC} shaped filters with 0.01 roll-off factor, generating an output signal at 8 samples/symbol. The shaped signals are then sent to I/Q \Gls{MZM} models for electro-optical conversion. The $2\,\mathrm{N_{m}}$ optical signals are then launched into the transmission fiber model with strong mode coupling. The fiber is modeled using the multisection scheme presented in \cite{ho2011mode} for 100 spans and 5,000 km total length. The channel consists of 1,000 frequency bins spread over 240 GHz (note that the simulation bandwidth is 30 GHz times 8 samples per symbol, yielding 240 GHz). The resolution of the channel in frequency domain is adjusted by replicating channel matrices between simulated frequency bins. 
    The group delay standard deviation is set to 3.1 ps/$\sqrt{\mathrm{km}}$ \cite{hayashi2017record}. The \gls{MDL}/\gls{MDG} of the link is controlled by a per-amplifier \gls{MDG} standard deviation, $\sigma_{g}$.
    After propagation, the received signals are converted from the optical to the electrical domain by the receiver front-end model. No phase noise has been considered for the simulations. The electric signals are down-sampled to two samples per symbol and fed into the \gls{MIMO} equalizer for source separation and equalization. 12$\times$12 MIMO equalization is carried out by 144 finite impulse response filters with 100 taps each, updated by a fully supervised least mean squares (LMS) algorithm. The \gls{MDG} standard deviation $\sigma_{\mathrm{mdg}}$ is computed at each frequency of the MIMO transfer function and averaged across the signal band.
    MDL/MDG estimation is performed without and with correction factor over the DSP-estimated eigenvalues. 
    \begin{figure}[t!]
        \captionsetup[subfigure]{labelformat=empty}
        \centering
        \subfloat[\label{subfig:curvassigmavssigmaNOCORR}]{%
            \includegraphics[width=0.99\linewidth]{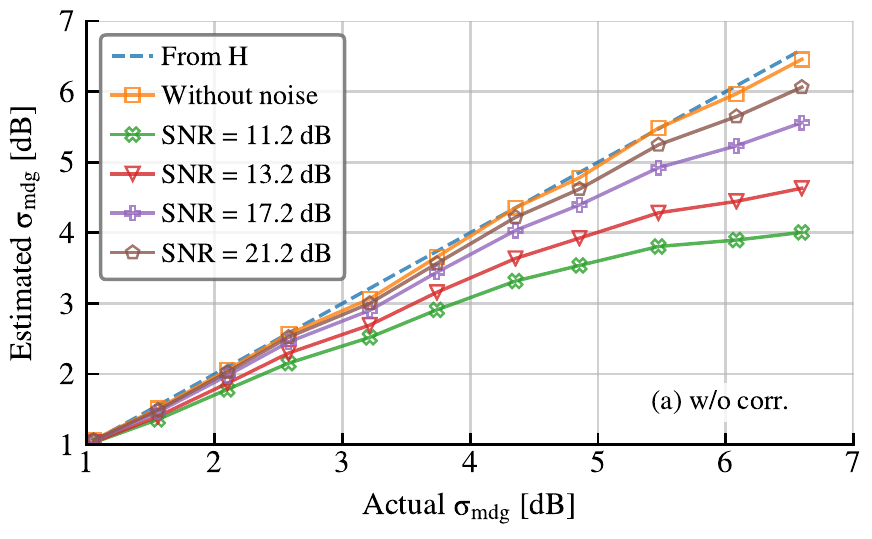}}
        \\ 
        \subfloat[\label{subfig:curvassigmavssigmaCORR}]{%
            \includegraphics[width=0.99\linewidth]{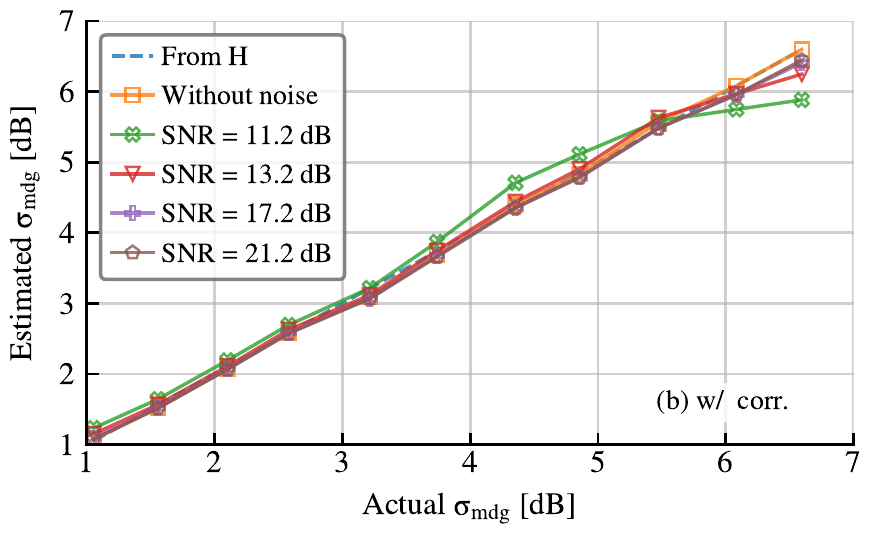}}
        \vspace*{-3mm}
        \caption{MDG standard deviation $\sigma_{\mathrm{mdg}}$ estimated from the equalizer coefficients, $\mathbf{W}^{-1}_{\mathrm{MMSE}}$, versus actual $\sigma_{\mathrm{mdg}}$, at different SNRs. (a) Without correction. (b) Correction by the positive solution of (\ref{Eq:roots}).}
        \label{fig:sigmavssigma}
    \end{figure}
    \cref{subfig:curvassigmavssigmaNOCORR} shows $\sigma_{\mathrm{mdg}}$ estimated by the coefficients of the MMSE equalizer as a function of the actual value without correction.
    In absence of noise, $\sigma_{\mathrm{mdg}}$ estimated from the equalizer coefficients tracks the actual $\sigma_{\mathrm{mdg}}$ with negligible error. As the SNR decreases, the estimation error increases for higher values of $\sigma_{\mathrm{mdg}}$, underestimating the actual \gls{MDL}/\gls{MDG}. 
    \cref{subfig:curvassigmavssigmaCORR} shows that DSP-based \gls{MDL}/\gls{MDG} estimation can be significantly improved using the positive correction factor derived in (\ref{Eq:roots}), resulting in a small residual error in the investigated range of $\sigma_{\mathrm{mdg}}$, even for the lowest SNR evaluated.

    \section{Experimental setup for the 3-mode transmission link}
    \begin{figure*}[t!]  
        \centering
        \includegraphics[width=0.96\textwidth]{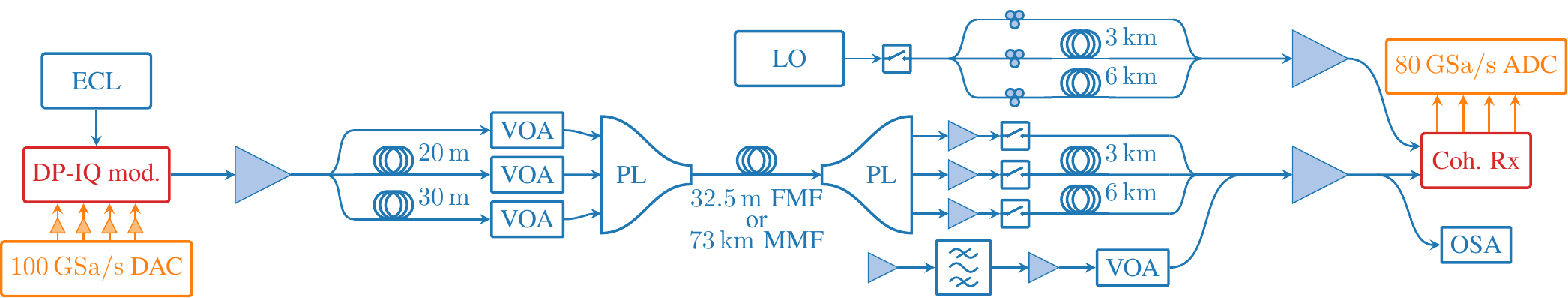}%
        \caption{Experimental setup for MDL/MDG emulation in short-reach 3-mode transmission with polarization multiplexing. The transmitter generates 16-QAM symbols at 25 GBd, which are subsequently split and delayed to create the input tributaries for the \gls{PL}. \glspl{VOA} are placed between the delay fibers and the \gls{PL} inputs in order to emulate MDL/MDG. The multi-mode signal is transmitted over either \SI{32.5}{\meter} of 4-LP FMF or over \SI{73}{\km} of \gls{MMF} \cite{sillard201650}. At the receiver, a \gls{TDMSDM} scheme is employed, and a noise loading stage is used to vary the \gls{OSNR}. MDL/MDG estimation is performed in DSP based on the MIMO MMSE transfer function.}%
        \label{fig:expsetup}
    \end{figure*}
    \label{sec:setup}
    The experimental setup used for MDL/MDG emulation is depicted in \cref{fig:expsetup}. At the transmitter, a \gls{PRBS} of $2^{16}$ polarization-multiplexed 16-QAM symbols is generated at 25 GBd. Pulse shaping at the transmitter is done using a \gls{RRC} filter with 0.01 roll-off factor. The pulse-shaped signal is converted to the analog domain by a 100 GSa/s \gls{DAC} followed by RF-amplifiers. The analog signal modulates the output of an \gls{ECL} operating at a frequency of \SI{193.4}{\tera\hertz} with a linewidth of 80 kHz using a dual-polarization in-phase and quadrature modulator. After optical modulation, the signal is amplified by an \gls{EDFA}, split and delayed by \SIlist{0; 20; 30}{\meter} to generate three decorrelated data streams that are passed through \glspl{VOA} and then multiplexed by a mode-selective \acrfull{PL} \cite{velazquez2018scaling}. The \glspl{VOA} allow the independent control of the launch powers in the 3 spatial modes \LP{01}, \LP{11a} and \LP{11b} to deliberately introduce MDL/MDG into the system. The output of the \gls{PL} is an FMF that supports 4 \gls{LP} mode groups. At the receiver side, a second PL is used as mode de-multiplexer.  Splicing the \gls{FMF} pigtails of the two photonic lanterns results in a  \SI{32.5}{\meter} link.
    We also evaluate a longer link by fusion splicing the FMF pigtails to a  \SI{73}{\km} fiber link consisting of 16 spools of \SI{50}{\micro\meter} core diameter graded-index \gls{MMF} \cite{sillard201650} with lengths varying from \SI{1.2}{\km} to \SI{8.9}{\km}.
     The receiver employs a time domain multiplexed (TDM)-SDM receiver \cite{van2014ultra} that delays two flows by 3 km and 6 km of standard single-mode fiber (SSMF) to reduce the required amount of the coherent receivers. After the TDM-SDM stage, a noise loading stage composed of two EDFAs, a wavelength selective switch (WSS) and a VOA is placed to vary the OSNR at the coherent receiver input. This noise-loading setup places a \SI{250}{\giga\hertz} wide noise-band around the \SI{193.4}{\tera\hertz} carrier. The average OSNR is measured by an \gls{OSA} after the last amplification stage. The SNR at the receiver input is computed as  $\mathrm{ SNR= OSNR \, (T_{s} \times 12.5 \,GHz)}$ where $\mathrm{T_{s} = 40 \, ps}$ is the symbol time \cite{essiambre2010capacity}. The noisy signal is amplified and converted from the optical to the electrical domain by the receiver front-end that integrates a second ECL as local oscillator (LO). The TDM electric signals are fed into 80 GSa/s analog-to-digital converters (ADC) to be digitized. In the DSP block, the TDM streams are parallelized and down-sampled to two samples per symbol. Next, in case of \SI{73}{\km} transmission, dispersion is digitally compensated and frequency offset is estimated and compensated for. The signal is matched-filtered by a RRC filter, and, finally, \gls{DD} equalization is applied.
    6$\times$6 MIMO equalization is carried out using a widely linear complex-valued adaptive equalizer, updated by a fully supervised DD-LMS algorithm \cite{da2016widely}.
 
    \begin{table}[!t]
        \caption{\label{configurations}VOA attenuation settings for different MDL/MDG emulation}
        \begin{center}
            \begin{tabular}{cccc}
                \toprule
                \multirow{2}{*}{\textbf{Case}} & \multicolumn{3}{c}{\textbf{Attenuation sweep}}       \\ \cmidrule{2-4}
                & \textbf{\LP{01}} & \textbf{\LP{11a}} & \textbf{\LP{11b}} \\ \midrule
                \textbf{1}                     & Decreases & Constant at $5$ dB & $5$ dB to $17$ dB     \\
                \textbf{2}                     & $5$ dB to $17$ dB & Constant at $5$ dB & Decreases   \\
                \textbf{3}                     & Decreases & $5$ dB to $17$ dB & $6$ dB to $18$ dB    \\
                \textbf{4}                     & Decreases & $5$ dB to $17$ dB & $5$ dB to $17$ dB     \\ \bottomrule
            \end{tabular}

        \end{center}
        \label{Tab:MDLcases}
    \end{table}
   
    \section{Experimental results for the 3-mode transmission link}
    \subsection{MDL/MDG estimation without noise loading}
\begin{figure*}[!ht]
        \centering
        \includegraphics[width=0.5\linewidth]{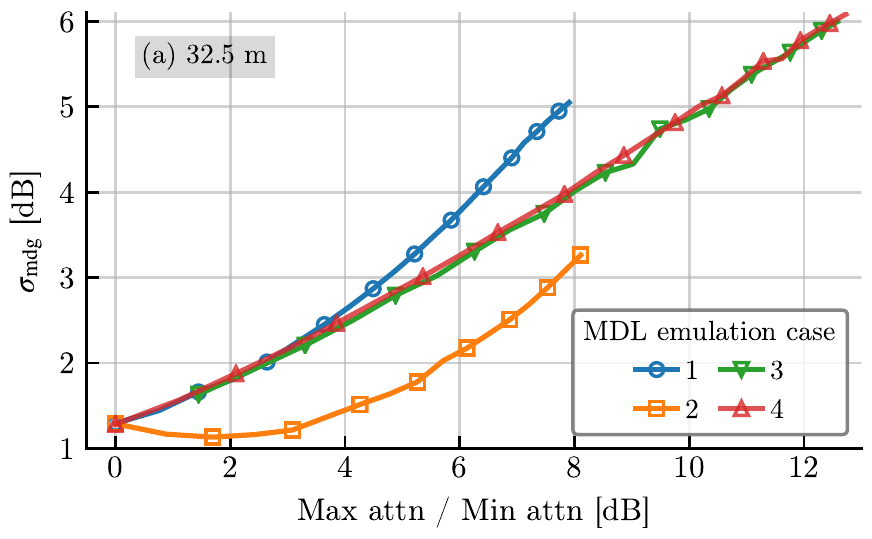}%
        \includegraphics[width=0.5\linewidth]{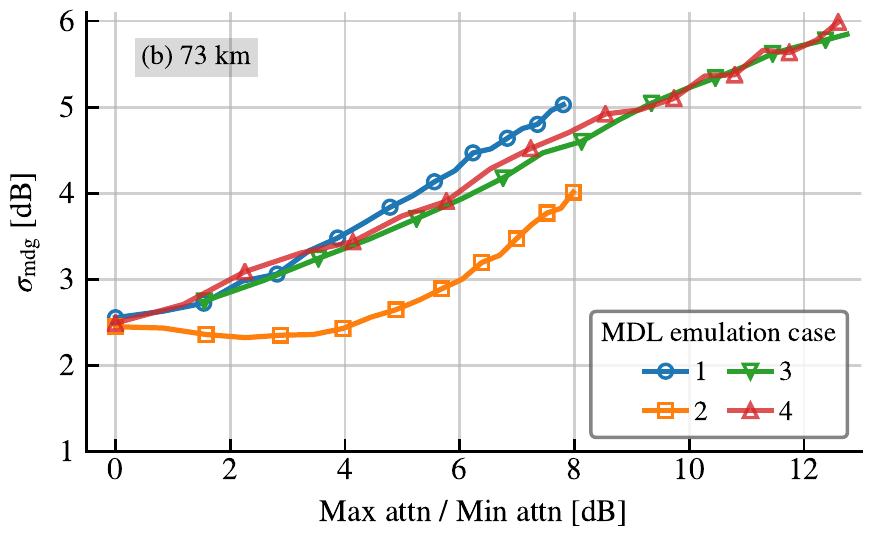}%
        \caption{MDL/MDG standard deviation $\sigma_{\mathrm{mdg}}$ versus attenuation ratio at different MDL/MDG emulation cases, estimated by DSP. (a) Transmission over \SI{32.5}{\m} without noise loading. (b) Transmission over \SI{73}{\km} without noise loading.}%
        \label{fig:mdl_vs_att}
    \end{figure*}

    The three \glspl{VOA} employed to vary the input power for the \LP{01}, \LP{11a}, and \LP{11b} ports of the \gls{PL} provide an attenuation range from 0~dB to 25~dB for an applied voltage between \SI{0}{\volt} and \SI{5}{\volt}. Since the relation between the applied voltage and the resulting attenuation is not linear, each individual \gls{VOA} is calibrated by scanning the applied voltage and measuring the attenuation. This data is used to generate a \gls{LUT}, which is interpolated to achieve an arbitrary attenuation.
    At first, the capability of the \glspl{VOA} to emulate the presence of MDL/MDG in the experimental setup is evaluated. At 0 dB attenuation, for \SI{32.5}{\m} transmission, the launch powers are $0.55$ dBm, $-0.15$ dBm and $-0.15$ dBm for \LP{01}, \LP{11a} and \LP{11b} ports of the \gls{PL}, respectively. For transmission over \SI{73}{\km}, the launch powers are measured to be $12.5$ dBm, $12.1$ dBm and $12.4$ dBm for \LP{01}, \LP{11a} and \LP{11b}, respectively. In order to keep the total launch power constant at $-4.9$ dBm for \SI{32.5}{\m} transmission and $7.3$ dBm for \SI{73}{\km} transmission, the 3 \glspl{VOA} are initialized in $5$ dB attenuation to attenuate or de-attenuate the signals according to the configurations defined in \cref{Tab:MDLcases}.
    In case 1, the \LP{11b} mode is gradually attenuated, while the attenuation of the \LP{01} mode decreases in such a way that the total launch power is constant. For case 2, the \LP{01} mode is attenuated instead of the \LP{11b} mode. In cases 3 and 4, the \LP{11a} and \LP{11b} modes are simultaneously attenuated, while the attenuation over the \LP{01} mode decreases.
    The induced MDL/MDG is estimated after DSP from the MIMO transfer function and averaged over 5 different captures. \cref{fig:mdl_vs_att} shows $\sigma_{\mathrm{mdg}}$ as a function of the ratio between the maximum and minimum attenuation for the four cases of \cref{Tab:MDLcases}.
    \begin{figure*}[!t]
        \captionsetup[subfigure]{labelformat=empty}
        \centering
        \subfloat[\label{subfig:powersweepNONOISE_b2b}]{%
            \includegraphics[width=0.483\linewidth]{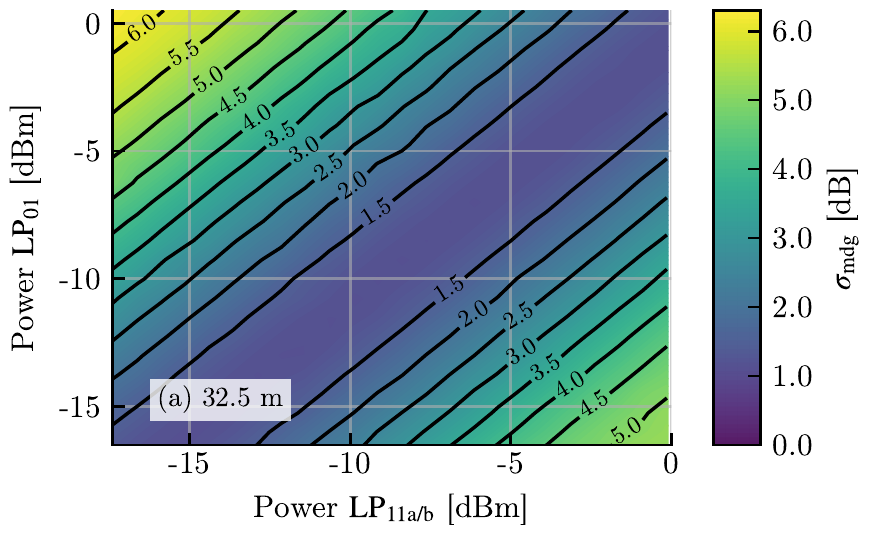}
            
            }%
        \hfill
        \subfloat[\label{subfig:powersweepNONOISE_73km}]{%
            \includegraphics[width=0.483\linewidth]{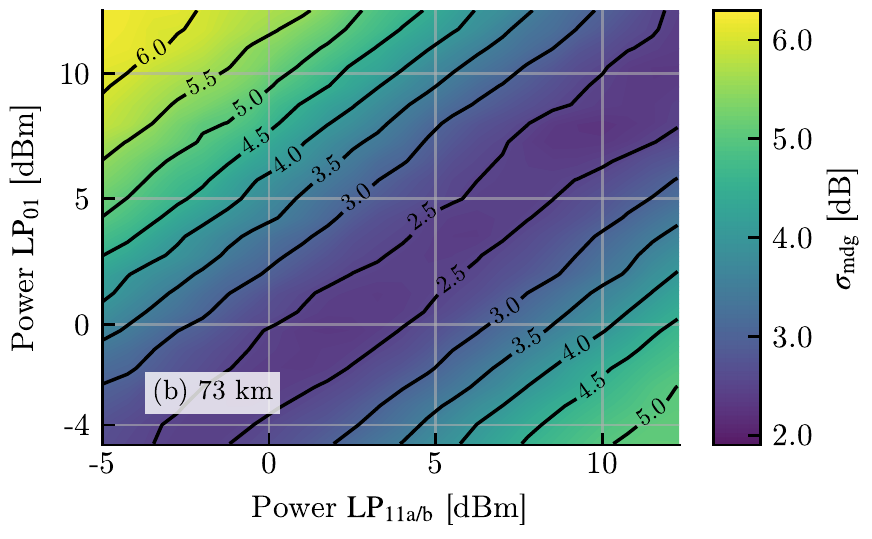}}%
        \caption{MDL/MDG standard deviation $\sigma_{\mathrm{mdg}}$ estimated by DSP without noise loading, as a function of the launch powers in the \LP{01} and \LP{11} modes. (a) \SI{32.5}{\m} transmission with an intrinsic SNR = 38.5 dB. (b) \SI{73}{\km} transmission  with an intrinsic SNR = 37.1 dB.}%
        \label{fig:powersweepNONOISE}
    \end{figure*}
      The different attenuation scenarios are used to emulate the system MDL/MDG. From \cref{fig:mdl_vs_att}a, at an attenuation ratio of 0 dB, the system has an inherent $\sigma_{\mathrm{mdg}} = 1.3$ dB for \SI{32.5}{\m} transmission, coming from the different launch powers, imperfections of the optical splitters and different insertion losses of the VOAs. In \cref{fig:mdl_vs_att}b, for \SI{73}{\km} transmission, a higher inherent $\sigma_{\mathrm{mdg}} = 2.6$ dB is measured as a consequence of the splices connecting the 16 spools. In general, for both \SI{32.5}{\m} and \SI{73}{\km} transmission, the higher the ratio between attenuations, the higher the induced MDL/MDG. In case 2, at low attenuation ratios, the induced MDL/MDG decreases slightly before turning into an increasing curve. Such behavior comes from the fact that, in this configuration, the strongest mode, \LP{01}, is more attenuated than modes \LP{11a} and \LP{11b}, compensating for the inherent launch power differences at 0 dB attenuation. Using cases 3 and 4, $\sigma_{\mathrm{mdg}}$ achieves up to $6$ dB as a consequence of the simultaneous attenuation of both \LP{11a} and \LP{11b} modes.
    
    Using the \glspl{VOA} to experimentally emulate MDL/MDG, Figs. \ref{fig:powersweepNONOISE}a and \ref{fig:powersweepNONOISE}b show the $\sigma_{\mathrm{mdg}}$ estimated by DSP as a function of the launch powers in the \LP{01} and \LP{11} modes achieved by sweeping the attenuation from 0 dB to 17 dB in the 3 \glspl{VOA} for \SI{32.5}{\m} and \SI{73}{\km} transmission, respectively.
    In the absence of noise loading, the SNR obtained from the OSNR measured optically by the OSA is 38.5 dB for transmission over \SI{32.5}{\m} and 37.1 dB for transmission over \SI{73}{\km}.
    From the contour plot in \cref{subfig:powersweepNONOISE_b2b}, $\sigma_{\mathrm{mdg}}$ increases from the middle of the grid towards the top left corner and the bottom right corner, where the difference between the launch powers is maximized. On the contrary, the region encompassing the diagonal between the bottom left corner and the top right corner presents low $\sigma_{\mathrm{mdg}}$ as a consequence of the high similarity between the launch powers. In \cref{subfig:powersweepNONOISE_73km}, for \SI{73}{\km} transmission, $\sigma_{\mathrm{mdg}}$ also increases in the direction in which the difference between the launch powers is maximized, achieving up to 6.3 dB. Over the region where the launch powers are similar, $\sigma_{\mathrm{mdg}}$ remains low around 2 dB.

    \subsection{MDL/MDG estimation with noise loading}
    \label{subsection:32.573kmmetersresults}
     We analyze the influence of noise on MDL/MDG estimation by loading noise to the optical transmission setup and calculating the estimation error $\sigma_{\mathrm{mdg}}^{\mathrm{err}}$, defined as the difference in dB between $\sigma_{\mathrm{mdg}}$, estimated in the setup without noise loading, and $\sigma_{\mathrm{mdg}}^{\textrm{nl}}$, estimated with noise loading ($\sigma_{\mathrm{mdg}}^{\mathrm{err}} = \sigma_{\mathrm{mdg}} - \sigma_{\mathrm{mdg}}^{\textrm{nl}}$).
     
\begin{figure*}[!t]
    \vspace*{-5mm}
        \captionsetup[subfigure]{labelformat=empty}
        \centering
        \subfloat[\label{subfig:powersweepERROR20dB}]{%
            \includegraphics[width=0.483\linewidth]{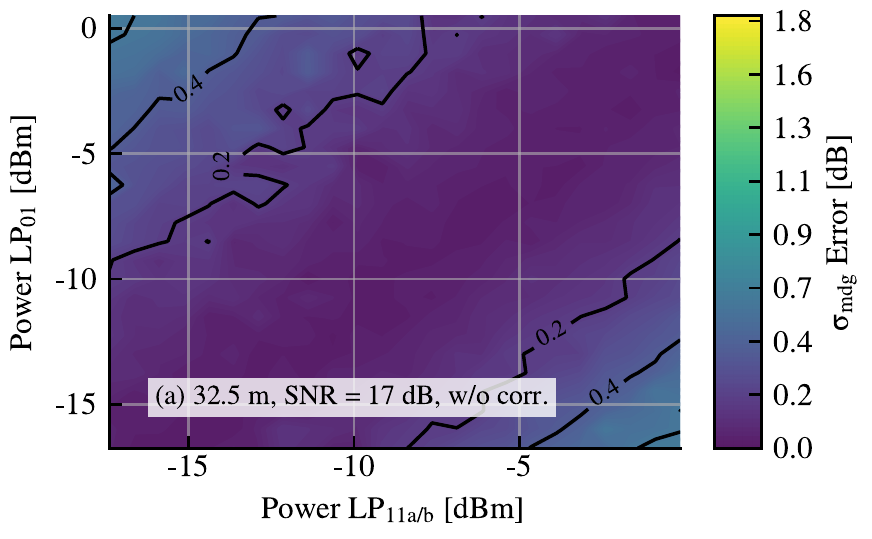}
            \label{fig:aaaa}}
        \hfill
        \subfloat[\label{subfig:powersweepERROR15dB}]{%
            \includegraphics[width=0.483\linewidth]{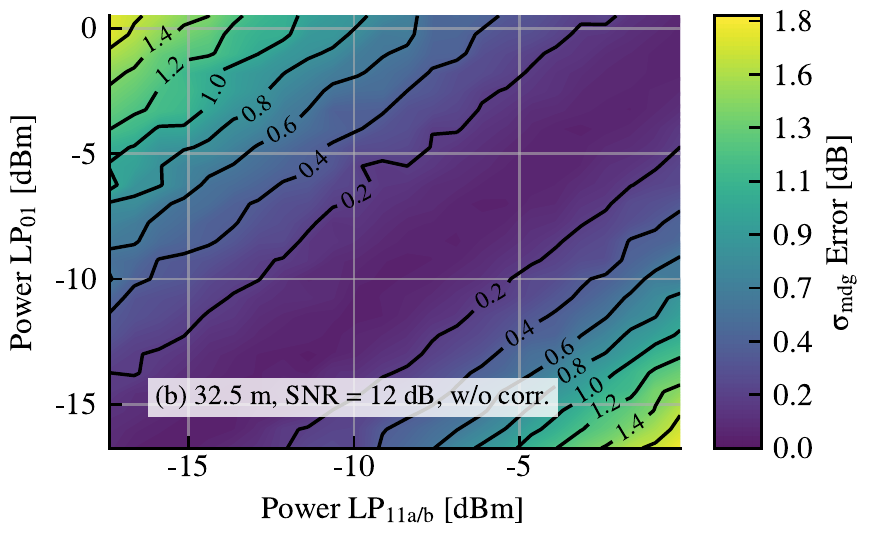}}
        \\ \vspace*{-8mm}
        \subfloat[\label{subfig:powersweepERROR20dBCORR}]{%
            \includegraphics[width=0.483\linewidth]{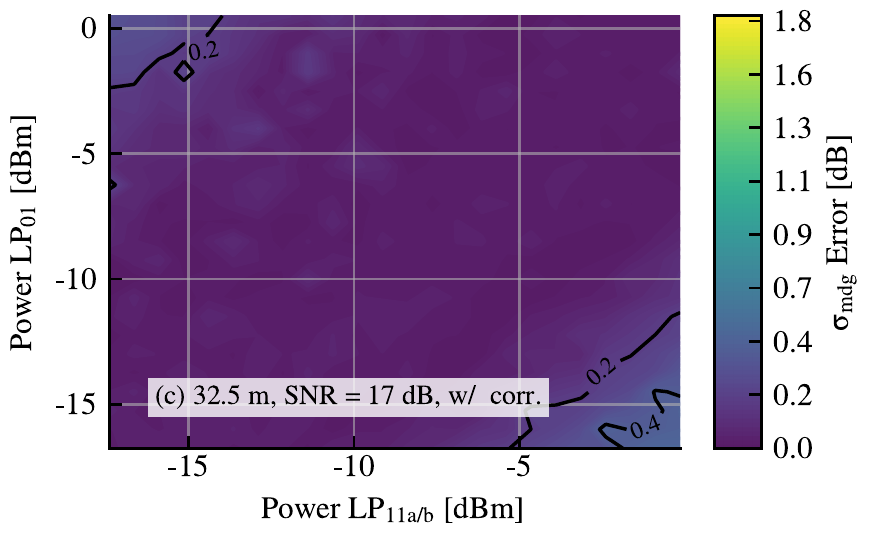}}
        \hfill
        \subfloat[\label{subfig:powersweepERROR15dBCORR}]{%
            \includegraphics[width=0.483\linewidth]{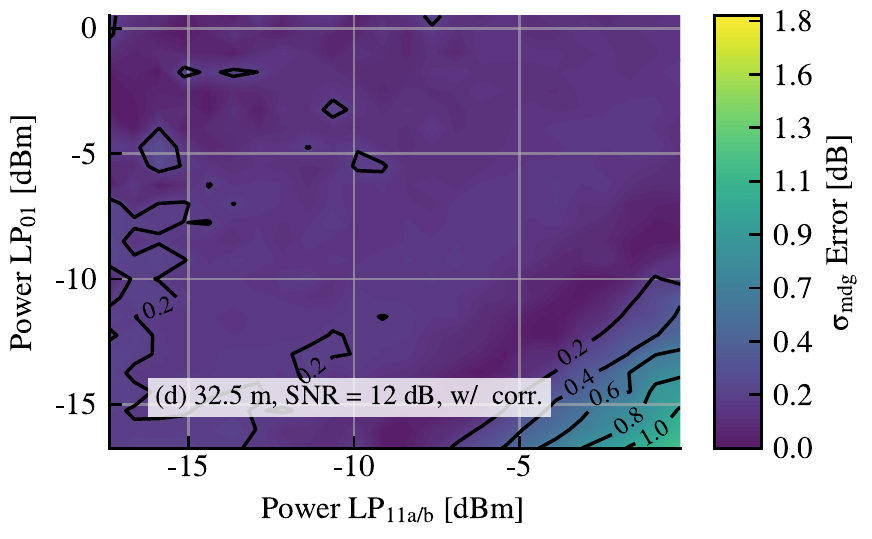}}
        \caption{Estimation error $\sigma_{\mathrm{mdg}}^{\mathrm{err}}= \sigma_{\mathrm{mdg}} - \sigma_{\mathrm{mdg}}^{\textrm{nl}}$, calculated as the difference in dB between the MDL/MDG standard deviation $\sigma_{\mathrm{mdg}}$ estimated by DSP without noise loading, and  $\sigma_{\mathrm{mdg}}^{\textrm{nl}}$ estimated by DSP with noise loading. In the figure, $\sigma_{\mathrm{mdg}}^{\mathrm{err}}$ is shown as a function of the power launched in the \LP{01} and \LP{11} modes for \SI{32.5}{\m} transmission. (a) Without correction at SNR = 17 dB. (b) Without correction at SNR = 12 dB. (c) Correction by the positive solution of (\ref{Eq:roots}) at SNR = 17 dB. (d) Correction by the positive solution of (\ref{Eq:roots}) at SNR = 12 dB.}
        \label{fig:powersweepERRORandCORR32}
    \end{figure*}

     Figs. \ref{subfig:powersweepERROR20dB} and \ref{subfig:powersweepERROR15dB} show $\sigma_{\mathrm{mdg}}^{\mathrm{err}}$ for the  \SI{32.5}{\m} transmission link, at SNR = 17 dB and SNR = 12 dB.
    As expected from the simulation results, $\sigma_{\mathrm{mdg}}^{\mathrm{err}}$ for SNR = 17 dB (up to 0.6 dB) is significantly lower than for SNR = 12 dB (up to 1.8 dB).
    The estimation error after correction is shown in Figs.  \ref{subfig:powersweepERROR20dBCORR} and \ref{subfig:powersweepERROR15dBCORR} for SNR = 17 dB and SNR = 12 dB. The correction factor given by the positive solution of (\ref{Eq:roots}) significantly enhances the estimation process over most of the grid, remaining only a small residual error in the high MDL/MDG regime for both SNRs. 
    
\begin{figure*}[!t]
        \vspace*{-5mm}
        \captionsetup[subfigure]{labelformat=empty}
        \centering
        \subfloat[\label{subfig:powersweepERROR20dB_73km}]{%
            \includegraphics[width=0.483\linewidth]{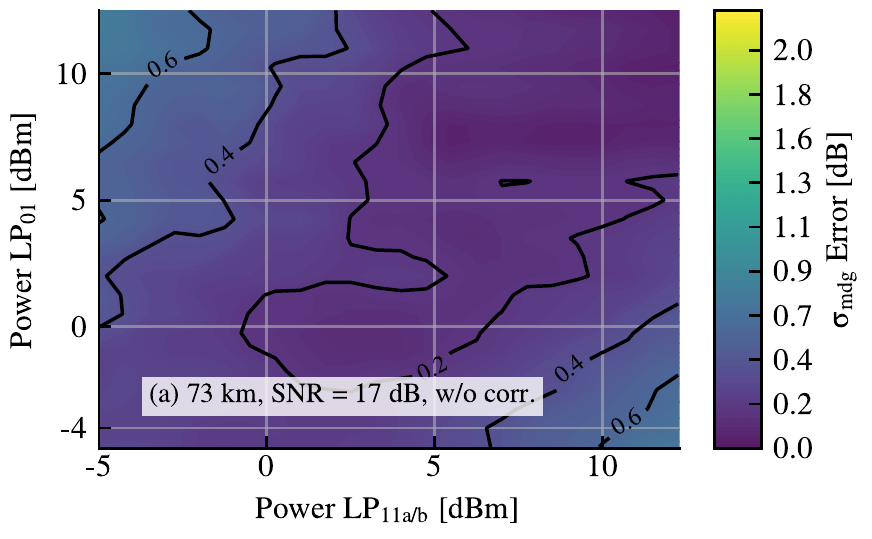}}
        \hfill
        \subfloat[\label{subfig:powersweepERROR15dB_73km}]{%
            \includegraphics[width=0.483\linewidth]{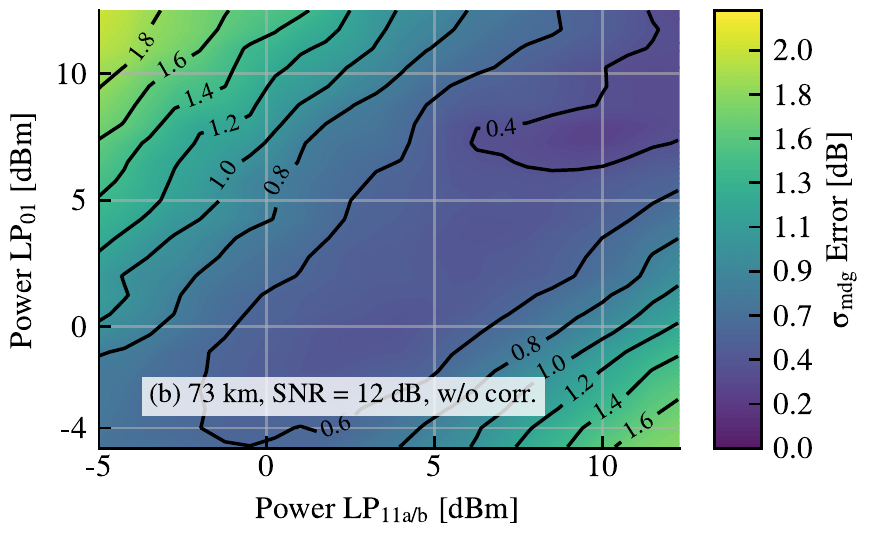}}
        \\ \vspace*{-8mm}
        \subfloat[\label{subfig:powersweepERROR20dBCORR_73km}]{%
            \includegraphics[width=0.483\linewidth]{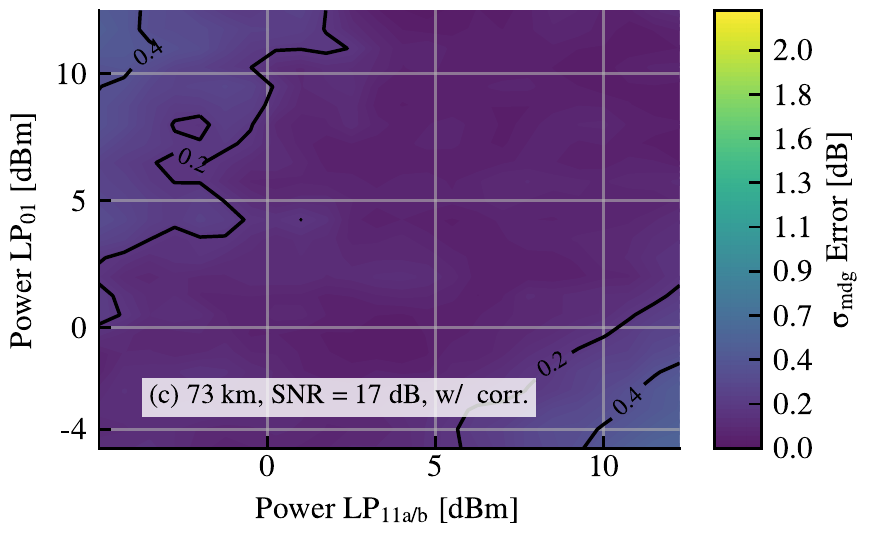}}
        \hfill
        \subfloat[\label{subfig:powersweepERROR15dBCORR_73km}]{%
            \includegraphics[width=0.483\linewidth]{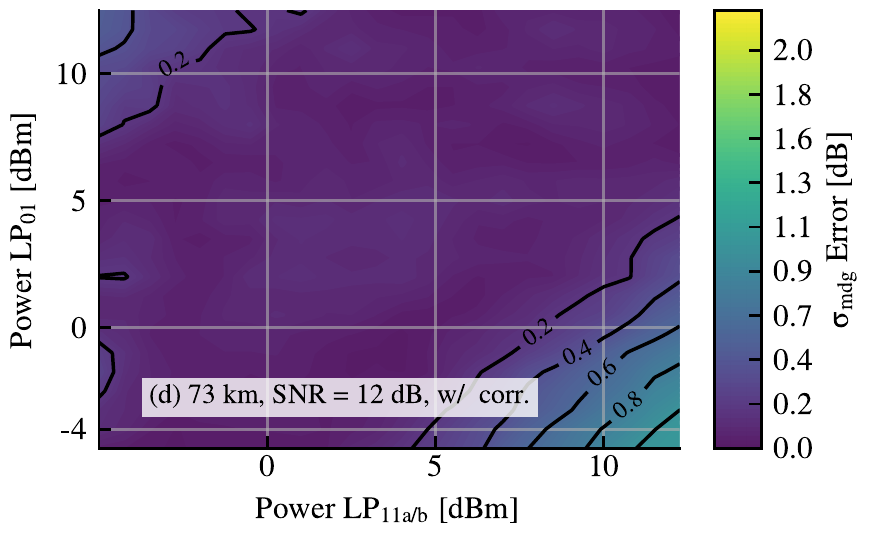}}
        \caption{Estimation error $\sigma_{\mathrm{mdg}}^{\mathrm{err}}= \sigma_{\mathrm{mdg}} - \sigma_{\mathrm{mdg}}^{\textrm{nl}}$, calculated as the difference in dB between the MDL/MDG standard deviation $\sigma_{\mathrm{mdg}}$ estimated by DSP without noise loading, and  $\sigma_{\mathrm{mdg}}^{\textrm{nl}}$ estimated by DSP with noise loading. In the figure, $\sigma_{\mathrm{mdg}}^{\mathrm{err}}$ is shown as a function of the power launched in the \LP{01} and \LP{11} modes for \SI{73}{\km} transmission. (a) Without correction at SNR = 17 dB. (b) Without correction at SNR = 12 dB. (c) Correction by the positive solution of (\ref{Eq:roots}) at SNR = 17 dB. (d) Correction by the positive solution of (\ref{Eq:roots}) at SNR = 12 dB.}
        \label{fig:powersweepERRORandCORR73}
    \end{figure*}

    The effect of noise on $\sigma_{\mathrm{mdg}}^{\mathrm{err}}$ for \SI{73}{\km} transmission is presented in Figs. \ref{subfig:powersweepERROR20dB_73km} and \ref{subfig:powersweepERROR15dB_73km}.
    The estimation error at SNR = 17 dB achieves up to 0.7 dB, while the estimation error at SNR = 12 dB reaches up to 2.2 dB. 
    The estimation error after correction is shown in Figs. \ref{subfig:powersweepERROR20dBCORR_73km} and \ref{subfig:powersweepERROR15dBCORR_73km} for SNR = 17 dB and SNR = 12 dB, respectively. As observed in the \SI{32.5}{\m} link, the correction factor significantly reduces the estimation error, remaining only a small residual error in the high MDL/MDG regime for both SNRs.
    
\begin{figure*}[t]
        \captionsetup[subfigure]{labelformat=empty}
        \centering
        \subfloat[\label{subfig:mdl_vs_osnr_b2b}]{%
            \includegraphics[width=0.483\linewidth]{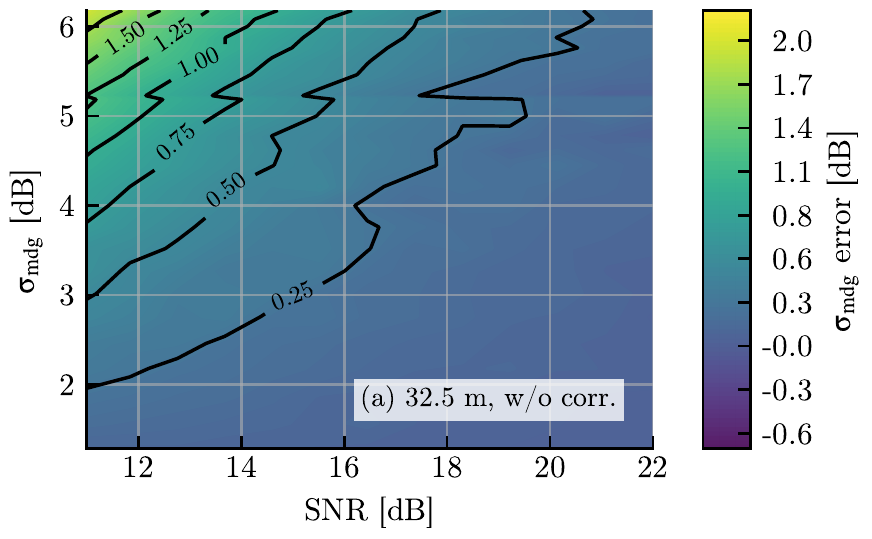}}
        \hfill
        \subfloat[\label{subfig:mdl_vs_osnr_b2b_corr}]{%
            \includegraphics[width=0.483\linewidth]{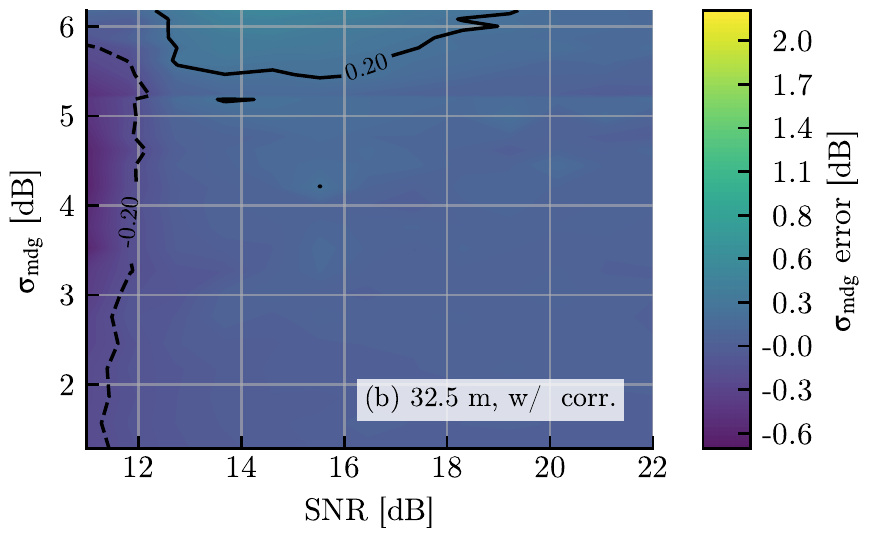}}
        \\
        \subfloat[\label{subfig:mdl_vs_osnr_73km}]{%
            \includegraphics[width=0.483\linewidth]{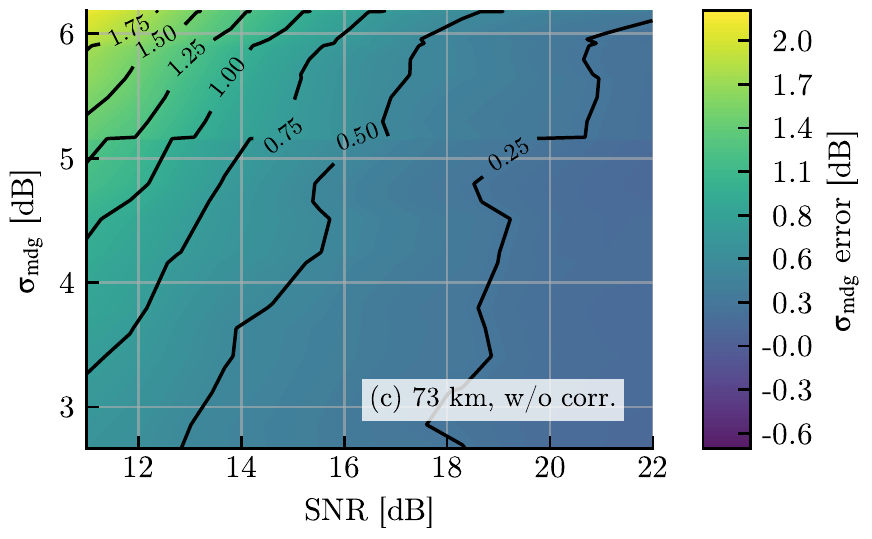}}
        \hfill
        \subfloat[\label{subfig:mdl_vs_osnr_73km_corr}]{%
            \includegraphics[width=0.483\linewidth]{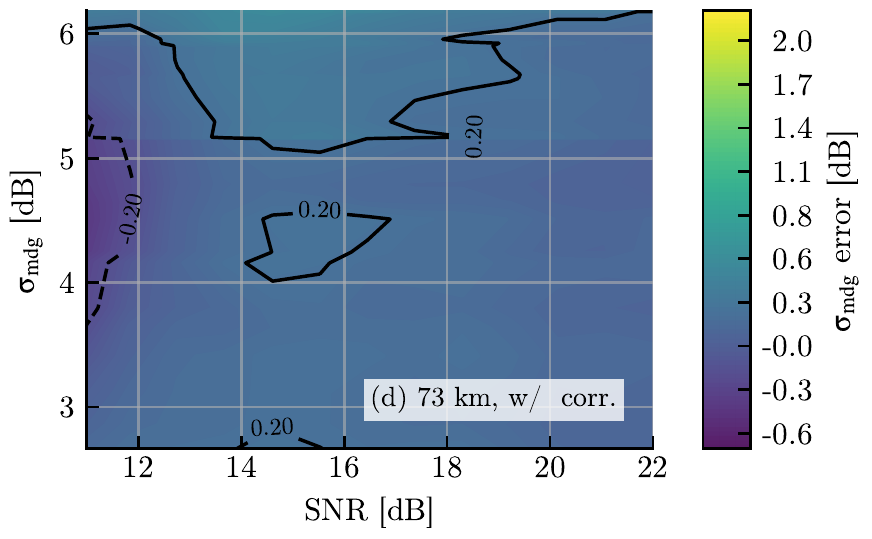}}
   
        \caption{Estimation error $\sigma_{\mathrm{mdg}}^{\mathrm{err}}= \sigma_{\mathrm{mdg}} - \sigma_{\mathrm{mdg}}^{\textrm{nl}}$, calculated as the difference in dB between the MDL/MDG standard deviation $\sigma_{\mathrm{mdg}}$ estimated by DSP without noise loading, and $\sigma_{\mathrm{mdg}}^{\textrm{nl}}$ estimated by DSP with noise loading. In the figure, $\sigma_{\mathrm{mdg}}^{\mathrm{err}}$ is shown as a function of $\sigma_{\mathrm{mdg}}$ and the SNR. (a) \SI{32.5}{\m} transmission without correction. (b) \SI{32.5}{\km} transmission with correction by the positive solution of (\ref{Eq:roots}). (c) \SI{73}{\km} transmission without correction.  (d) \SI{73}{\km} transmission with correction by the positive solution of (\ref{Eq:roots}).}
        \label{fig:mdl_vs_osnr}
    \end{figure*}

    Next, we sweep both the attenuation of the VOAs and the noise power at the receiver input. Figs. \ref{fig:mdl_vs_osnr}a and \ref{fig:mdl_vs_osnr}b show $\sigma_{\mathrm{mdg}}^{\mathrm{err}}$ as a function of $\sigma_{\mathrm{mdg}}$ and SNR without and with correction of the DSP-estimated eigenvalues for \SI{32.5}{\m} transmission.
    In \cref{subfig:mdl_vs_osnr_b2b}, without correction, the estimation error achieves up to 1.8 dB for $\sigma_{\mathrm{mdg}}> 6$ dB and SNR $<$ 12 dB. At an SNR = 12 dB, the estimation error varies from 0.25 dB to 1.7 dB across the range $2 \: \mathrm{dB}< \sigma_{\mathrm{mdg}}<6$ dB. As the SNR increases, the estimation error decreases progressively. At an SNR $>$ 20.5 dB, the estimation error is less than 0.25 dB for all ranges of $\sigma_{\mathrm{mdg}}$.
    The correction factor applied over the DSP-estimated eigenvalues enhances the estimation across all ranges of $\sigma_{\mathrm{mdg}}$ and SNR evaluated in \cref{subfig:mdl_vs_osnr_b2b_corr}. Here, a residual error of 0.2 dB is achieved for the high MDL/MDG regime. For SNR $<$ 12 dB, the residual error is negative as a consequence of the over-correction of the eigenvalues that results in an estimated $\sigma_{\mathrm{mdg}}$ higher than the actual $\sigma_{\mathrm{mdg}}$.
    The results for \SI{73}{\km} transmission are shown in Figs. \ref{fig:mdl_vs_osnr}c and \ref{fig:mdl_vs_osnr}d.
    In \cref{subfig:mdl_vs_osnr_73km}, without correction, the estimation error achieves up to 2 dB for $\sigma_{\mathrm{mdg}}> 6$ dB and SNR $<$ 12 dB. At an SNR = 12 dB, the estimation error varies from 0.75 dB to 1.75 dB across the range $3.5 \: \mathrm{dB}< \sigma_{\mathrm{mdg}}<6$ dB. An estimation error less than 0.25 dB for all values of $\sigma_{\mathrm{mdg}}$ is obtained for SNR values higher than 22 dB.
    The correction factor significantly reduces the estimation error, as shown in \cref{subfig:mdl_vs_osnr_73km_corr}. In this case, only a residual error of 0.2 dB is observed in certain regions of the grid. For SNR $<$ 12 dB and $3.7 \: \mathrm{dB}< \sigma_{\mathrm{mdg}}<5.2 \: \mathrm{dB}$, there is a negative residual error of $- 0.2$ dB as a consequence of over-correction.
    


    \glsreset{SDM}
    \glsreset{MDL}
    \glsreset{MDG}
    \glsreset{SNR}
    \glsreset{MIMO}
    \glsreset{MMSE}

    \section{Conclusion}
    In \gls{SDM} systems with coupled channels, the achievable channel capacity and transmission distance are fundamentally constrained by noise, \gls{MDL} and \gls{MDG}.
    MDL/MDG not only reduce the average capacity but can cause outages. MDL/MDG estimation carried out by coherent receivers is a useful tool for link assessment and troubleshooting. In this paper, we show that MDL/MDG estimation carried out directly from the dynamic equalizer coefficients is prone to errors in regimes of low \gls{SNR} and high MDL/MDG. Using the transfer function of an equalizer based on the \gls{MMSE} criterion, we calculate a correction factor that improves the estimation process in moderate levels of MDG/MDL and \glspl{SNR}. We validate the correction method by Monte-Carlo simulations of a 6-mode long-haul coupled transmission processed by a 12$\times$12 dynamic equalizer. Moreover, we experimentally validate the correction factor in a 3-mode transmission link using the coefficients of a 6$\times$6 dynamic equalizer for both \SI{32.5}{\m} and \SI{73}{\km} transmission. The simulations and experiments confirm the applicability of the method in practical transmission scenarios.


    \ifCLASSOPTIONcaptionsoff
    \newpage
    \fi



%
    \bibliographystyle{IEEEtran}
    \bibliography{refR}
%








\end{document}